\documentclass[10pt, a4paper]{article}

\usepackage{url}
\usepackage[round]{natbib}
\usepackage{authblk}
\usepackage{graphicx}
\usepackage{pdfpages}

\title{A Blueprint for Public Engagement Appraisal: Supporting Research Careers}
\author[1,2]{Josh Borrow}
\author[1]{Pedro Russo}
\affil[1]{Leiden Observatory (russo@strw.leidenuniv.nl)}
\affil[2]{Durham University (joshua.borrow@durham.ac.uk)}

\date{\today}

\begin{document}

\maketitle

\begin{abstract}
Time spent performing public engagement is severely undervalued by research institutions around the globe. In this article we present one possible system that could be implemented to provide researchers with career recognition for performing this vital work. The framework utilises the supervision system that is already in place at many research institutions, whereby senior researchers mentor their junior colleagues. This would encourage more researchers to engage with the public, as well as increasing the quality of this engagement.
\end{abstract}

\clearpage
\section*{Executive Summary}

In this report, we outline a method for providing career recognition to researchers who engage with the public. 

The anti-engagement atmosphere, along with a systemic undervaluation of public engagement in research institutions today is discouraging participation in science outreach at all career stages. Young researchers in particular are actively dissuaded from “wasting their time” on such activities by more senior colleagues; a stark contrast to the third aim of research institutes around the globe, which is to facilitate knowledge transfer to the public.

This attitude is unsurprising, as currently most career development in research institutions is tied directly to research and, to a smaller extent, teaching. This not only includes training opportunities, but also promotion and pay progression. If a researcher would like to engage with the public, they are prevented from doing so. They will have less time to devote to their research and teaching duties, which are directly connected to their salary. This is, perhaps, why one of the most often stated reasons for not engaging with the public is a lack of time; it is not so much a lack of time itself, but a lack of time that researchers are willing to sacrifice for their own personal values.

It is important to note that we are not suggesting that each and every researcher engages with the public. Simply that time spent performing good public engagement work is treated similarly to time spent producing good research.

The solution to the anti-engagement atmosphere and the need to promote more public engagement proposed here is focused not only on appraisal but also supporting researchers in their initiatives. The format for the solution is a supervision system similar to the one employed to mentor junior researchers by their more senior colleagues; researchers will have meetings with their supervisor to discuss their public engagement work and receive support as well as to summatively evaluate their work. From these appraisals, each researcher can build their own public engagement portfolio similarly to how they build a research portfolio with a list of publications.

This portfolio should then be utilised by the research institution as well as other bodies when evaluating the researcher's performance, with time spent performing good public engagement work weighted equally alongside time spent producing good research.

We make the following recommendations to research institutions:
\begin{itemize}
\item Implement an evaluation system for public engagement
\item Use these evaluations to build a public engagement portfolio for both individual researchers and the institute as a whole
\item Promote public engagement within the institute
\item Have an experienced public engagement researcher present when discussing performance lead pay, promotion, etc.
\item Utilise the aforementioned portfolio to weight time spent performing research and public engagement equally, if they are of the same standard as judged by the member above.
\end{itemize}
\clearpage

\section{Introduction}
\label{sec:intro}

Public engagement is often described as the `third aim' of research institutions behind research and teaching. However, this is not reflected in their career development programmes or the atmosphere within these institutions \citep{view_mat, bio_phys}. This leads to public engagement initiatives being undertaken by a small, self-selected group of people who either enjoy outreach or feel that it is part of their moral duty as a researcher, rather than seeing it as part of their career \citep{ind_ner}.

There is research  supporting the hypothesis that one of the main reasons researchers do not undertake public engagement is that they do not receive career recognition for doing so. For example, some of the researchers interviewed by \citet{car_wat} were forced to abandon their public engagement work for fear of being left behind in their career.  Available research suggesting that researchers are not motivated to engage with the public by the prospect of it contributing to their career development \citep{view_bes} is often limited by small sample sizes and large amounts of self-selection bias and as such cannot be used to draw meaningful conclusions.

This attitude of not including participation in public engagement in career development is especially clear when looking at job descriptions. As of Friday the 3\textsuperscript{rd} of July 2015, out of 34 research positions on the American Astronomical Society job register (from PhD to permanent professor), only one has a single mention of outreach in it. The institution noted that they employ people to do outreach. There was no mention of this being part of the researcher's job, which is in severe conflict with the supposed `third aim'. It is particularly surprising to see this when in the same job descriptions there were mentions of, among other things, the quality of public transport in the town where the institution is located.

Other research suggests an anti-engagement atmosphere within research institutions discourages public engagement work. Geneticists who were interviewed by \citet{view_mat} were concerned about discouragement from senior colleagues. Physicists and Biologists interviewed by \citet{bio_phys} suggested that engaging with the public can even be detrimental to career development. The anti-engagement atmosphere that exists within many research institutions is unsurprising; those who proliferate it are trying to protect their colleagues from throwing their career away. 

All of this is happening at a time when some researchers, research institutions and funding bodies are calling for increased public engagement \citep{surv_niel}. There are several good reasons to engage the public, from all of these viewpoints. Firstly, engaging the public can be rewarding. This is perhaps the most underrated reason for undertaking public engagement from an institution's point of view. Secondly, public engagement work can raise the profile of an institution with the public, leading to higher application rates for undergraduate and postgraduate courses and increased brand awareness. Articles that appear in the media are more likely to be cited, for example \citep{pub_kie}. Thirdly, research performed by the crowdfunding platform, One Percent Club, \citep{csr_len} shows that employees that undertake socially responsible work (e.g. public outreach at a research institution) are more engaged and productive, as well as having more pride in the organisation they work for. The argument has also been made that undertaking public engagement work helps researchers with teaching and even research itself. This hypothesis is supported by a study of French researchers that showed that ones that engage with the public more are also more active academically, however a causal link was not investigated \citep{eng_jen}.

If our aim is to engage the public, then we need to motivate researchers to do so. For this to work effectively, we also need to foster a pro-engagement atmosphere. Both of these issues can be remedied by providing more career development to researchers who engage with the public.
\section{Evaluation}
\label{sec:eval}

There has been a considerable amount of work over the past few decades towards developing evaluation techniques for public engagement initiatives. However, these techniques usually involve very specific quantitative and qualitative methods which require a large amount of time and effort to implement \citep{writ_tsa, mj_sha, ane_wei}. These metrics can be useful for large-scale informal learning institutions such as science centres, museums but for individual scientists working in public engagement they represent an considerable investment of resources and time.

Even within these large-scale institutions, evaluation with metrics is often incredibly difficult due to the large number of factors that need to be considered \citep{box_nere}. The efforts to evaluate these programmes are also often misguided, which is a waste of both time and money \citep{prob_jen, evi_king}. Long-term learning outcomes from a single interaction in particular are nearly impossible to measure.

\subsection{What is evaluation?}

Evaluation collects information that can be analysed to find a project's weaknesses, strengths and effects. Successful evaluation takes place throughout a project, allowing it to be moulded to fit changing needs as well as keeping it on track when this is unnecessary. We can consider evaluation to be split into three categories:
\begin{itemize}
\item \emph{Initial} evaluation occurs at the beginning of the project. Here it is useful to focus on the 5 Ws (who, what where, when and why).
\item \emph{Interim} evaluation is the longest stage, taking place throughout the planning and implementation of the project. A focus on keeping the goals set at the outset in focus is important.
\item \emph{Summative} evaluation occurs at the end of the project. This focuses on questions such as ``did we reach our target audience?" and looks at if the goals and actions defined throughout the project were realised.
\end{itemize}

\subsection{Why evaluate?}

Evaluation can be loosely defined by the application of the following points \citep{eval_iya}:
\begin{itemize}
\item Find out if the target audience for the project was reached
\item Identify areas where the project was lacking or inefficient
\item Discover unexpected results
\item Motivate the group by showing the outcomes of their effort
\item Obtain suggestions for improvement
\item Determine if the objectives of the project have been achieved
\item Consider how the project is going to evolve and find ways to facilitate this.
\end{itemize}

Evaluation is then a process for determining the evolution of a project, including the final form. This can be used to prove what the project did well and how it can be improved in the future.

Successfully evaluating a project has several advantages for the public, the researcher and their institution. This results from improved time management due to increased goal focus, effectively allowing researchers to do more with less time by avoiding wasting effort on avenues that could have been identified as unproductive \citep{goal_sch}. A successful evaluation can also be used for appraisal of the project, leading to the possible inclusion of public engagement into the current career recognition system in place at research institutions.

\subsection{Evaluation guides}

Evaluation is an exceptionally challenging subject. Over evaluation and under evaluation are present in a large number of projects. Under evaluation leads to future projects suffering from the same underlying issues as past ones. A lack of evaluation can be at any stage, and a lack of initial evaluation can be particularly damaging. This can lead to confusion about the outcome of the project and what actions to take \citep{fettu_arc}. For example, a lack of initial evaluation for the America COMPETES Act lead to a possible loss of community partnerships \citep{nasa_yee}. Over evaluation can also be an issue, preventing researchers from focusing on their actual public engagement work. Below are a number of guides that can be used to find a successful balance.

\begin{itemize}
\item ``The evaluation of science-based organisations' communication programs" \citep{eval_met} is a good, short introduction to evaluating public engagement initiatives. The authors outline a seven-step approach borrowing from work done previously for public health programs.
\item ``A Guide for Successfully Evaluating Science Engagement Events" \citep{eval_gam} is another short guide on evaluating small-scale events. It has a section on the motivation behind evaluation that is of particular note.
\item ``Public Participation Methods: A Framework for Evaluation" \citep{eval_rowe} provides a more in-depth overview of evaluating public engagement initiatives where, in particular, the views of the public are collected.
\item ``Highlighting the value of evidence-based evaluation: pushing back on demands for impact" \citep{evi_king} contains a case-study of evaluation focusing on a narrative approach rather than using difficult to measure metrics.
\item ``Evaluation: Practical Guidelines" by Research Councils UK \citep{eval_rcu} is a comprehensive guide for evaluating public engagement activities.
\end{itemize}
\section{Goal Setting}
\label{sec:goals}

When planning projects, it is useful to set goals. These are ideals that are to be completed by the conclusion of the project, or even may describe the project itself. The process of setting goals allows individuals to focus on what is important about their project as well as realise areas that would represent wasted effort.

Goal setting has been used in recent years in many contexts, from business to mental health \citep{schut_goals}, but what is most relevant here is its use in coaching. Goal setting can be used as a motivational tool as well as an evaluation framework \citep{gs_locke_2}.

Using goal setting as a framework allows for easy evaluation. If applied correctly, it allows for a self-sustaining system of evaluation and application, where evaluation motivates goal setting, which in turn motivates the next round of evaluation. This will be described in more detail in section \ref{sec:frame} but, in short, allows for solid evaluation with minimal time input such that researchers can attain the benefits outlined in section \ref{sec:eval}.

\subsection{Values, goals and actions}

It is instructive to split what are colloquially known as `goals' into three categories:
\begin{itemize}
\item \emph{Values} are the highest order. These are general overview ideas that guide the individual and are generally more abstract, for example ``Public engagement activities should be evaluated better".
\item \emph{Goals} are medium order and are motivated by the above values. These are more concrete concepts, for example ``Create an evaluation framework for public talks".
\item \emph{Actions} are the lowest order and are the individual items that are used to realise the above goals. For example ``Read papers on jargon use and public engagement".
\end{itemize}

\subsection{Setting SMART goals}

In section \ref{sec:eval}, we stressed the importance of evaluation.
As such, it makes sense to create a goal-setting framework that is compatible with easy evaluation.

A popular framework for goal setting is SMART \citep{smart_dor}. This means to set goals that are:
\begin{itemize}
\item \emph{Specific} goals facilitate group work in particular by preventing overlap.

\item \emph{Measurable} goals are fit for evaluation. Note that this does not imply that there needs to be a specific metric associated with each goal.

\item \emph{Assignable} goals also facilitate group work by ensuring that each individual member of the team can understand their role.

\item \emph{Realistic} goals allows for researchers to stay motivated. We must acknowledge that most researchers will not see public engagement as their primary focus and as such it is nonsensical to expect everyone to undertake a large-scale project.

\item \emph{Time-bound} goals are key for motivation as well as aiding with evaluation.
\end{itemize}
This framework is exceptionally useful, helping to set well targeted goals and actions that can be easily implemented and evaluated.

\subsection{Evaluation of goals}

One of the main reasons to use a goal-lead system is that it sets out items that can be evaluated individually; it is considerably easier to evaluate a project in small actions than it is to build a complete picture from the outset.

\subsection{Setting the best goals}

The following discussion borrows heavily from the work presented by \citet{gs_locke}. This paper is a short, useful introduction to goal theory and should be read by anyone interested in implementing a framework based on goal setting.

Goals should be specific (as outlined above), but also the bar should be set reasonably high (i.e. the goals should be difficult) to increase motivation and performance. This may mean researchers may have to stretch themselves and go outside of their comfort zone to achieve their best performance for time input. These high goals are also accompanied with a higher satisfaction, as is to be expected.

Having goals that align with peers also represents a group performance increase, again as is to be expected. If the whole group works on a project collaboratively, or individual projects that are similar, this will foster a better atmosphere and allow peers to provide considerably better feedback based on their own experiences.

It is extremely important that goals align with personal values. This may mean researchers choose activities that are tailored especially to them. For example, researchers with young children may prefer to work in primary schools educating youngsters on the basics in their field.

\subsection{Goals in practice}

A useful analogy to understand how values, goals and actions fit together is to consider climbing a mountain peak. Many different people may have different motivations for doing this. One example is that they would like to be recognised for being a brave, strong person. Another is that they want to prove to themselves that they can climb the tallest mountain in the world. The summit is clearly the goal here, it is what you aim for in practice. Actions take the form of the planning to climb the mountain (booking flights etc.), and actually making the arduous slog up the slope.

This highlights the importance of setting realistic goals. Whilst it is unlikely that one would die on the way to realising their goal of, for example, performing a public talk, it is very easy for ambitious researchers to over-stretch themselves. This leads to poor performance in the activity and everyone involved feeling dissatisfied.

It also highlights the importance of focusing on values and goals, rather than actions. It is unlikely that anyone would enjoy the process of booking a flight, or spending many hours trying to procure a visa. This is the same in public engagement. It may not be all that fun arguing with a venue because they messed up the date that you were supposed to be giving your talk on, but just like climbing the mountain, once you reach the summit the satisfaction will be worth the effort.
\section{Supporting career development with public engagement}
\label{sec:frame}

Currently, most career development in research institutions is tied directly to research and, to a some extent, teaching. This not only includes training and development opportunities, but also promotion and pay progression. This means that if a researcher would like to engage with the public, they are discouraged from doing so as they will have less time to devote to their research and teaching duties that are directly connected to their salary. This is, perhaps, why one of the most often stated reasons for not engaging with the public is a lack of time \citep{bio_phys, view_mat}.

To increase engagement, there needs to be a network in place that supports practitioners in their efforts and provides them with appraisals that can be utilised when career progression opportunities arise. These appraisals can be built into a public engagement ``portfolio" by the researcher similarly to how a list of publications and previous research appraisals are compiled. To implement such a system will not be without difficulties, so to aid with the transition we suggest that the pre-existing is used as a foundation. This system, as well as increasing the quality of interaction between researchers and the public, will allow for recognition of work that is historically undervalued.

Within this system, the focus should be on producing good quality public engagement initiatives with as little administrative work as possible. We are trying to encourage participation in public engagement, not paperwork. 

The institution must be completely committed to ensuring that such a system is effective. It is of little use to compile a large quantity of data on an individual's public engagement efforts and then ignore it when discussing performance-lead pay. Note that we are not suggesting that all researchers be forced to engage with the public. We are proposing that time spent performing good public engagement work is treated similarly to time spent producing good research.

\subsection{Outline}

\begin{figure}[h!]
	\centering
    \includegraphics[width=\textwidth]{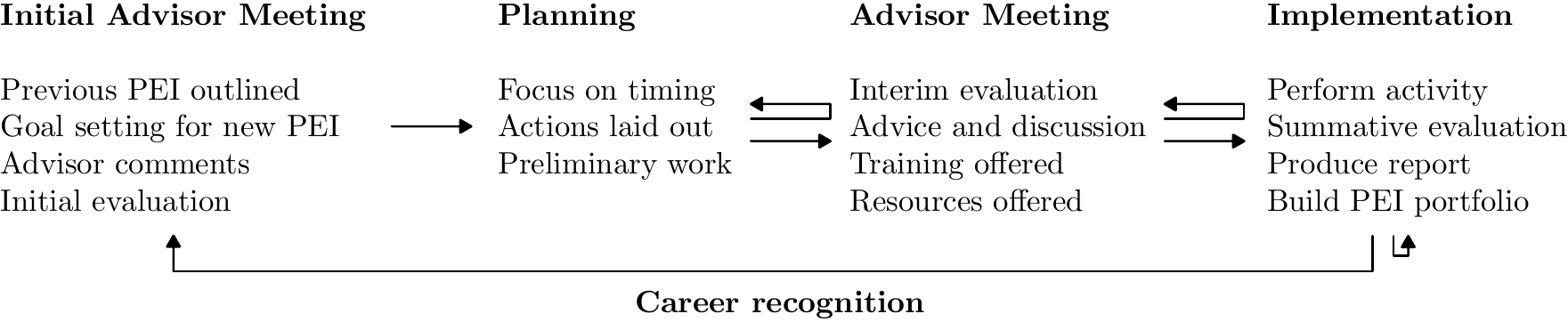}
    \caption{General overview of the goal-setting evaluation system. Evaluation should take place throughout the process, and each stage may need to occur more than once depending on the scale of the public engagement initiative (PEI).}
    \label{fig:gs}
\end{figure}

As we have indicated above, there is a clear need for a public engagement evaluation system within research institutions. This evaluation system should serve to support researchers as well as help them appraise their work.

The general idea is that each researcher will have a supervisor who they will meet with several times a year to discuss their engagement projects, evaluating as they go along. This supervisor, similarly to a research supervisor, will also give the researcher advice and help them find relevant training opportunities to further their outreach career. At the conclusion of a public engagement initiative, the researcher will produce a summative evaluation which is then appraised by the supervisor. This appraisal and evaluation can then be placed into a public engagement `portfolio' by the researcher, along with other media materials such as a video recording of a public talk, for example. The portfolio can then be published on their personal webpage next to their list of publications, as well as being made available in other formats. When a career development opportunity presents itself, this portfolio can be produced, similarly to how publications can be used to show research output, to provide evidence of good previous public engagement work.

The key to success here is continual evaluation. Without this, there can not be a solid appraisal at the conclusion of the project, as well as the project suffering as a result. The evaluation can be guided by using a pre-existing evaluation framework, including those already in use at the research institution or one that was presented in section \ref{sec:eval}.

\subsection{Initial meeting}

Before a discussion of new projects can begin, the researcher should outline previous public engagement work. If there has been no meeting to discuss appraisals of this previous work, then it should take place here. This is of particular importance if there has been an effort to summatively evaluate previous work by the researcher and there has been no time to schedule a meeting.

The next part of the meeting should begin with setting an overarching value that aligns with both the research institution's goals and the researcher's personal values. From this value, goals can be set for projects to undertake that will satisfy it. These goals should be at least guided by the principles of SMART goal setting, as outlined in section \ref{sec:goals}.

Once goals have been set, the researcher and their supervisor should now discuss the 5 Ws, i.e.
\begin{itemize}
	\item \emph{Who}: Who is the target audience for this project? How are we going to make sure that we reach them? Can we collaborate with anyone?
    \item \emph{What}: What are we trying to accomplish here? Is there an ulterior motive that we could satisfy better by using a different activity?
    \item \emph{When}: When will the event take place?
    \item \emph{Where}: Where will the event take place?
    \item \emph{Why}: Why this goal? What do I want to get out of it? What will the target audience want to get out of this interaction?
\end{itemize}
These 5 Ws should be used to guide the rest of the project. In particular, the target audience is important. Everything should be pitched correctly towards this audience, including branding, media and invitations. There is little point inviting four year old children to your in-depth discussion of black holes. Normally, \emph{how} is also included in this list, however here that is dedicated to the next three stages of the system.

\subsection{Planning}

Here, actions are laid out and performed. These actions, as with the goals, should be SMART. Each action should be laid out with a specific time goal; this will act as both a motivational tool and aid with evaluation.

At each stage of the project, the 5 Ws laid out at the beginning should be considered. If the project has changed its scope away from these, then perhaps a re-evaluation is required.

Once the actions are planned, work can begin. This will involve, for example, making a powerpoint presentation for a public talk.

\subsection{Meeting(s)}

Depending on the scale of the project, more than one secondary meeting may need to take place. This meeting is an opportunity to perform serious interim evaluation, to ensure that the project is still on track. It is important here to consider the budget and schedule of the project before any small issues end up out of control. There should be a serious conversation about the projects direction and possibilities to make it even more successful. 

Here the researcher and supervisor should consider the 5 Ws together, and discuss if they are all working together in the correct way. For example, if the venue has been decided to be a university lecture theatre at 3 pm, this is incompatible with a target audience of young professionals.

Extra training and resources should be requested by the researcher or suggested by the supervisor, and funding for putting this in place should be allocated.

Further actions for the future should be considered by both the researcher and supervisor and laid out here. A plan for summative evaluation should be discussed.

\subsection{Implementation}

After the completion of the activity, a summative evaluation should take place. This evaluation should determine if the goals were satisfied, and how the project could be better implemented in the future.

In this evaluation, there should also be a record of the narrative of the event. The researcher's views on how successful the event was should be laid out here, with easy to measure metrics such as attendance.

There should be no effort to measure `impact'. Whilst this approach does have downfalls, as we have discussed above there is no place for this kind of evaluation in small-scale public engagement. It is simply too expensive, both in money and time investment.

There should, however, be an effort to measure simple metrics such as approximate audience composition for a public talk such that it can be determined if the correct audience was reached. A simple questionnaire handed to a small sample of the audience is enough to gauge opinion, along with the researcher's narrative. Of course, this is never going to be fully representative, but it can provide useful information on how similar projects should be performed in the future. Again, it is important to note that we are trying to aid researchers in public engagement, not turn them into sociologists.

A report should be produced that can be added to the researcher's public engagement portfolio, with comments from their supervisor. This portfolio can then be used when career development opportunities present themselves, rather than relying on research metrics alone.
\section{Case study}
\label{sec:case}

All of the materials that are presented below using the case study are available online for free to facilitate implementation of this system. Please see in the references section, under \citet{mat_borr}, for more details.

The case study below outlines a possible implementation of the appraisal scheme presented in section \ref{sec:frame}. This case study shows what we believe to be a reasonable level of evaluation; a small enough amount such that the researcher does not spend the majority of their time evaluating, but enough to see if the event has been successful or not. A highlight here is the lack of abstract metrics utilised; here there is no obtuse effort to measure `impact'. As we have discussed earlier, impacts such as long-term knowledge acquisition are impossible to measure even for large-scale informal learning institutions and efforts should not be made to measure them by individual researchers.

Once the event has been completed, along with a video of the public talk, these forms would be placed into the researcher's public engagement portfolio. This portfolio could be presented on the web, as part of their personal webpage and appearing alongside their list of publications. Having portfolios as part of personal webpages not only helps individual researchers, but raises the engagement profile of the institution as a whole. It also helps foster a pro-engagement atmosphere by normalising research as part of an individuals career, which contributes to promoting engagement as an `acceptable' use of researcher's time.



# PEI Initial Evaluation Form

**Date:** 2015-08-04

| | |
|---|---|
| **Name:** Jack Harkness | **Position:** Postdoc |

**Value:** Children should be inspired to undertake a "space career"

| | |
|---|---|
| **Goal:** Give talk on space careers at a local high school | **Goal:** Give a public talk on black holes |
| **Who:** High school students | **Who:** Everyone: mainly lower-level secondary school (11-14) but all should be welcome. |
| **What:** Talk and questions and answer session | **What:** Public lecture/talk about black holes with time for questions afterwards. |
| **Where:** Gallifrey high school | **Where:** Big lecture hall within university (C107) |
| **When:** September/October this year (before the Oxbridge UCAS deadline) | **When:** Half term around February. 7pm on a weekday to ensure parents can attend. |
| **Why:** To inspire more students to apply for space-related degree programmes and apprenticeships | **Why:** There is some misconception about black holes I would like to clear up. Also I hope to inspire attendees |
| **Barriers to overcome:** need to focus on things other than academic careers. Need to find time in the school day/convince school that the talk is worth giving. Need to be *interesting* and tailor to a high-school audience. Need to not dress too 'sciencey'. | **Barriers to overcome:** Convincing university to use room could be difficult. **Publicity** is extremely important here. Don't use too much jargon. Don't panic (I do not have that much lecturing experience). |
| **Initial actions:** Contact airbus about technical careers by Friday. Contact school by Friday. Contact admissions tutor (physics) by Friday about UCAS forms as to give good advice to the students. | **Initial actions:** Contact security/building managers about after hours use of C107 by Friday. Write some notes to base the talk around by next Friday. |
| **Plan for interim evaluation:** Check if the presentation is catering to the correct audience. See if I contacted the school with good time to spare. | **Plan for interim evaluation:** Ensure room booked. Ensure a lack of (unexplained jargon) in the presentation/talk. Have advisor look over posters and note down how many we've printed. |

**Advisor initial thoughts:** Good initial ideas. Jack seems to understand the need for publicity for the second goal and hopefully this will play out in his favour. Not overly ambitious which is good. No extra resources have been requested at this time for any projects.

| | |
|---|---|
| **Advisor:** David Tennant | **Signature:** |

# PEI Interim Evaluation Form   **Date:** 2015-08-04

| **Name:** Jack Harkness | **Position:** Postdoc |
|---|---|
| **Advisor name:** David Tennant | **Signature:** |

**Goal:** Give talk on space careers at a local high school

**Goal:** Give a public talk on black holes.

**Is the project on schedule? Why (not)?** Yes, I will be giving the talk on October 11th. This is working well because I contacted the school early.

**Is the project on schedule? Why (not)?** Yes, the room is booked for Feb. 12th at 7pm.

**Is the project on budget? Why (not)?** N/A

**Is the project on budget? Why (not)?** No budget needed so far. I would like to have catering provide tea/coffee at the event which we can arrange at a later time.

**Is the project satisfying the 5Ws laid out initially? If not, why? What can be changed?** Yes. We just need to ensure that I keep targeting the audience correctly.

**Is the project satisfying the 5Ws laid out initially? If not, why? What can be changed?** Yes, however we need to market the event successfully in January, perhaps fairly aggressively. Otherwise nobody will turn up.

**Advisor suggestions:** Keep the presentation short so that the children don't loose interest.

**Advisor suggestions:** Posters and e-mails should be sent to local schools before half term. Contact the secretaries ASAP about catering for the event.

**Extra training/resources requested:** No resources needed.

**Extra training/resources requested:** Money for T&C.

**Actions for the future (may be separate):** Give talk on the 11th. Buy a thunderbolt to VGA adaptor by the 10th October.

**Actions for the future (may be separate):** Contact secretaries today. Write notes by the 30th of October. Make powerpoint presentation by 1st December. Make promo material before 30th December. Send e-mails and posters to schools by the 15th January. Give talk.

# PEI Summative Evaluation  **Date:** 2015-08-04

| **Name:** Jack Harkness | **Position:** Postdoc |
|---|---|
| **Advisor name:** David Tennant | **Signature:** |

| **Goal:** Give a talk on space careers at a local high school | **Goal:** Give a public talk on black holes. |
|---|---|

**Was the project completed on time? Why (not)?** Yes, the presentation was completed.

**Was the project completed on time? Why (not)?** Yes.

**Was the project completed on budget? Why (not)?** There was no budget for this project.

**Was the project completed on budget? Why (not)?** Yes, the department paid for the tea and coffee and the room hire was provided for free.

**Did the project satisfying the 5Ws laid out initially? If not, why? What can be changed?** Yes, the talk was given to the students at the high school, and I think I inspired some of them to take up space careers.

**Did the project satisfying the 5Ws laid out initially? If not, why? What can be changed?** Yes, apart from the audience. There were a lot more older people there than I anticipated (50+), but that is fine also. There were still a lot of schoolchildren, which I think was down to my promotional efforts.

**Advisor comments:** Good project. Should run next year perhaps in a few more schools if he wants.

**Advisor comments:** Good promotion was the key to success. The talk was well delivered (I went along) and in my opinion was pitched at the correct level.

**Narrative of the event (did attendees enjoy etc.):** The students seemed to enjoy it, and a lot of them were very interested in the aerospace engineering apprenticeship at airbus. I gave them contact details and a few of them seemed very excited. I also have received e-mails from a couple of students asking for help with their UCAS forms which I am happy to give.

I don't think I managed to dispel the 'classical scientist' image, but that's just me. The teachers seemed appreciative for the talk, which is good. Feedback from the questionnaires I gave the students and teachers was overwhelmingly positive, but small sample.

**Narrative of the event (did attendees enjoy etc.):** 152 people came to the event, with about 30% of those schoolchildren, 40% 50+ and the rest were parents accompanying the children. From my observations of the crowd, and due to the fact that many people asked questions at the end, I think my talk was engaging and well enjoyed. A questionnaire did not seem appropriate here as it would have taken longer to sift through them than to deliver the talk. I would like the opportunity to give a similar talk again next year.

The talk raised the profile of the university, people were surprised by the strong research profile.


\section{Conclusion}
\label{sec:con}

It is clear that there is a serious undervaluation of public engagement activities by research institutions. Their efforts to promote public engagement as their third aim are well intentioned, but lacking in support. If these institutions are serious about public engagement, then they need support systems in place to aid researchers. They currently risk alienating their staff and the public, as well as damaging their reputation.

In addition to being supported, it is imperative that researchers are not forced into undertaking public engagement, rather, if they want to then it is possible to do so without feeling like they have to make career sacrifices.

When evaluating public engagement activities a focus on the overall narrative of the event, rather than considering specific metrics or searching for `impact', is more productive. This approach prevents researchers from undertaking large amounts of administrative work and allows for them to focus on engaging with the public. Of course, this is not a suggestion to disregard evaluation, but rather to acknowledge a different approach is more suitable.

If a system such as the one outlined in section \ref{sec:con} above is implemented, then this will also help foster a pro-engagement atmosphere. The system will raise the profile of public engagement and will prevent critics from using the phrase ``waste of time'' to describe it.

From all of this, we have some policy recommendations for research institutions wanting to realise their third aim:
\begin{itemize}
\item Implement an evaluation system for public engagement
\item Use these evaluations to build a public engagement portfolio for both individual researchers and the institute as a whole
\item Promote public engagement within the institute
\item Have an experienced public engagement researcher present when discussing performance lead pay, promotion, etc.
\item Utilise the aforementioned portfolio to weight time spent performing research and public engagement equally, if they are of the same standard as judged by the member above.
\end{itemize}

\section{Acknowlegements}

The authors would like to thank John Tobin and Silvia Toonen, as well as the rest of the organisational staff who allowed this research to take place as part of the the Leiden/ESA Astrophysics Program for Summer Students.

\bibliography{bib.bib}

\end{document}